\documentclass[prb,aps,twocolumn,amsfonts,showpacs]{revtex4}
\usepackage{epsfig}
\usepackage{psfrag}
\usepackage{color}

\def\be{\begin{equation}}
\def\ee{\end{equation}}
\def\bea{\begin{eqnarray}}
\def\eea{\end{eqnarray}}

\newcommand{\new}{\textcolor{black}}

\begin{document}

\title{\new{${\rm SU(N)}$ Evolution of a Frustrated Spin Ladder}}
\author{Miraculous J. Bhaseen and Alexei M. Tsvelik}
\affiliation{Department of Physics, Brookhaven National Laboratory,
  Upton, NY 11973-5000, USA} 
\date{\today}

\begin{abstract}
\new{Recent studies indicate that the weakly coupled, $J_\perp \ll J$, spin-$1/2$ Heisenberg antiferromagnet with next nearest neighbor frustration, $J_\times$, supports massive spinons for $J_\times=J_\perp/2$. The straightforward ${\rm SU(N)}$ generalization of the low energy ladder Hamiltonian  yields two independent ${\rm SU (N)}$ Thirring models  
with ${\rm N}-1$ multiplets of massive ``spinon'' excitations.} We \new{study the evolution of} the complete
set of low energy dynamical structure factors using form
factors. Those \new{corresponding to} the smooth (staggered) magnetizations are qualitatively different (the same) in the ${\rm N}=2$ and ${\rm N}>2$ cases.
The absence of single-particle peaks
\new{preserves} the notion of spinons stabilized by frustration.
In contrast to the ladder, we note that the ${\rm N}\rightarrow\infty$ limit of the four chain \new{model} is not a trivial free theory. 
\end{abstract}

\maketitle


\section{Introduction}
Frustrated quantum antiferromagnets are a source of considerable theoretical and experimental attention --- see for example reference \onlinecite{Frustrated:Lhuillier}. Their characteristics include enhanced classical ground state degeneracies and the suppression of long-range N{\'e}el order. In addition to their intrinsic interest, their prominance is fueled by the high-${\mathrm T_c}$ superconducting cuprates, where hole doping frustrates, and ultimately destroys  the
long-range N{\'e}el
order of the parent compounds --- see for example \onlinecite{Sachdev:Order}.
This motivates the quest for simple
models of frustrated quantum magnets, and a detailed understanding of their properties. 

Important examples include nearest neighbor antiferromagnets on frustrated lattices, such as the triangular,\cite{Collins:Triangular} pyrochlore, and
Kagom{\'e}\cite{Azaria:Kagome} lattices, and further neighbor models on
regular lattices. The second variety embraces frustrated chains\cite{White:Dimerization} and ladders,\cite{Allen:Frustrated,Nersesyan:Incommensurate,Allen:Fate} the planar pyrochlore,\cite{Palmer:Quantum,Fouet:Planar,Tchernyshyov:Bond} and the square lattice antiferromagnet with next-nearest neighbor interactions. Indeed, the
latter model was suggested by Anderson in his influential work\cite{Anderson:RVB} on
$\mathrm {La_2CuO_4}$, as a means to
realize his ``resonating-valence-bond'' (RVB) or
``quantum spin-liquid'' state. With isotropic nearest neighbor exchange, $J_1$, this is often referred to as the $J_1-J_2$ model --- for an introduction to spin-liquids see Chapter 6 of the book by Fradkin.\cite{Fradkin:Field} Other examples include multispin exchange models, and those of dimers.\cite{Moessner:RVB} Although enormous progress continues to be made, frustrated quantum magnetism remains theoretically challenging. In general one must resort to $1/{\mathrm S}$ or $1/{\mathrm N}$ expansions, numerical simulations,
or other approximation schemes --- see for example reference \onlinecite{Capriotti:Quantum}. 

\begin{figure}
\begin{center}
\noindent
\epsfysize=0.2\textwidth
\epsfbox{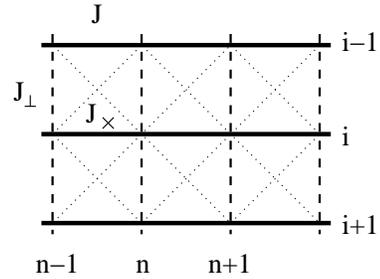}
\end{center}
\caption{\label{fig:nnnafm}2D Heisenberg antiferromagnet with next nearest neighbour frustration $J_\times\le J_\perp\ll J$. The strongly relevant interchain interaction between staggered magnetizations vanishes for $J_\times=J_\perp/2$ and renders deconfined spinons.}
\end{figure}

Building on the work of reference \onlinecite{Allen:Fate}, Nersesyan and Tsvelik have made considerable advances in the so-called confederate flag model.\cite{Nersesyan:Spinons}  This is an anisotropic version of the much studied $J_1-J_2$ model, in which the
nearest neighbor exchange has a strongly prefered chain direction --- see
figure \ref{fig:nnnafm}.  
The limit $J_\times\le J_\perp\ll J$ may be viewed as a collection of weakly coupled, but nevertheless interacting chains,  and field theory methods may be employed.
In general, the massless spinons of the spin-1/2 chain\cite{Faddeev:What} are confined by the interchain interactions. However, along the line, $J_\times=J_\perp/2$, massive spinons emerge in pairs, as the elementary spin excitations of the coupled system.\cite{Allen:Fate,Nersesyan:Spinons} In general they are neither bosons nor fermions, but have momentum dependent scattering.  There have been many speculations about the existence of such excitations in two-dimensional frustrated antiferromagnets, and their possible r{\^o}le in high-${\rm T_c}$. The developments of reference \onlinecite{Nersesyan:Spinons} deserve further investigation.

\new{In this paper we return to an ${\rm SU(N)}$ generalization of the ladder introduced in reference \onlinecite{Allen:Fate}.} Our motivation is twofold: firstly, the large-${\mathrm N}$ approach is known to miss qualitative features in this case,\cite{Allen:Fate} and we wish to track its evolution in detail. Large-${\mathrm N}$ results will be important in two-dimensions, and we hope to gain expertise in all the solvable cases. Secondly, we calculate the dynamical structure factors of the
staggered magnetizations. 
These involve correlation functions of {\it interacting} WZNW fields, and their evaluation beyond the ladder, is a highly challenging and open problem.\cite{Nersesyan:Spinons,Smirnov:Propagating}

The layout of this paper is as follows: in section \ref{model} we reacquaint the reader with the \new{spin-1/2} model, and it's mapping on to two different ``parity'' sectors. \cite{Allen:Fate,Nersesyan:Spinons} \new{We introduce the {\rm SU(N)} variant of the low energy action and comment on this choice of generalization. We emphasize that this treatment is not the same as replacing lattice spins by ${\rm SU(N)}$ generators;\cite{Lecheminant:Private}} we expand on this in appendix \ref{app:spin} where we comment on the connection between filling and ${\rm SU({\rm N})}$ ``spin'' representations. In section \ref{sect:dsf} we calculate the dynamical susceptibilities \new{corresponding to} the uniform and staggered magnetizations. We conclude in section \ref{sect:conc} with results for the four chain model.  In appendix \ref{NATM} we discuss in detail, the excitations, scattering and form factors of the $\mathrm{SU(N)}$ Thirring model. We hope this may be of some assitance to the unfamiliar reader.

\section{Model}
\label{model}
\begin{figure}
\begin{center}
\noindent
\epsfysize=0.08\textwidth
\psfrag{1}{${\rm I}$}
\psfrag{2}{${\rm II}$}
\epsfbox{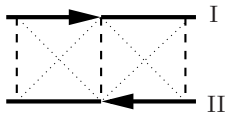}
\end{center}
\caption{\label{fig:plusparity} The ladder Hamiltonian is the sum of two independent ${\rm SU(N)}$ Thirring models: ${\mathcal H}={\mathcal H}_++{\mathcal H}_-$. In the sector of positive ``parity'' the even (odd) chains carry left (right) moving fields. The sector of negative parity is obtained by reversing the arrows.  Excitations of the ladder carry this $\pm$ index and may be produced in both sectors.}
\end{figure}

In this section we reacquaint the reader with the \new{spin-1/2} confederate flag model model, and it's mapping on to two different ``parity'' sectors;\cite{Allen:Fate,Nersesyan:Spinons} we shall specialize to the ladder in due course. Consider a Heisenberg antiferromagnet on a two-dimensional square lattice (of spacing $a_0$) with next nearest neighbor exchange interaction $0<J_\times\le J_\perp\ll J$ as depicted in figure \ref{fig:nnnafm}:
\bea
\label{NNNAFM}
H & = & \sum_{i=1}^L\sum_n\left[J{\bf S}_{i,n}\cdot{\bf S}_{i,n+1}+J_{\perp} {\bf S}_{i,n}\cdot{\bf S}_{i+1,n}\right.\nonumber\\
&& \hspace{0.4cm}+\left. J_{\times}\left( {\bf S}_{i,n} \cdot {\bf S}_{i+1,n+1}
+ {\bf S}_{i,n+1} \cdot {\bf S}_{i+1,n}\right)\right].
\eea
It is well established, that the low energy dynamics of a single \new{spin-1/2} (isotropic) Heisenberg  chain
\be
H_i^{\rm 1D}=\sum_n J{\bf S}_{i,n}\cdot{\bf S}_{i,n+1},
\ee
are described by  the \new{$\widehat{\rm su}(\rm 2)_1$} Wess--Zumino--Novikov--Witten (WZNW) model;\cite{Affleck:Critical,Affleck:Exact} for a review see \onlinecite{Affleck:FSCP,Tsvelik:boson}. This WZNW model has conserved currents ${\bf J}={\mathcal L}_\alpha^\dagger {\bf t}_{\alpha\beta} {\mathcal L}_\beta$ and ${\bf \bar J}={\mathcal R}_\alpha^\dagger {\bf t}_{\alpha\beta} {\mathcal R}_\beta$, which generate the \new{$\widehat{\rm su}(\rm 2)_1$} Kac--Moody current algebra, and the Hamiltonian density, $H=\int dx \,{\mathcal H}$, may be written in the following (Sugawara) form:
\be
\label{1dsug}
{\mathcal H}_i^{{\rm 1D}}={\mathcal N}\,\hbar v\,(:{\bf J}_i\cdot{\bf J}_i:+:{\bf \bar J}_i\cdot{\bf \bar J}_i:)+\cdots
\ee
Here $v$ is the spin velocity,  ${\mathcal N}$ is a  normalization constant, and the ellipsis stand for less relevant operators. We replace the perturbing lattice spin operators by their continuous, slowly varying, uniform and staggered components:
\be
\label{magstag}
{\bf S}_{i,n}/a_0\rightarrow {\bf S}_i(x)={\bf M}_i(x)+(-1)^n{\bf N}_i(x), 
\ee
where $x\equiv na_0$ measures the distance along chain $i$. Neglecting oscillatory and derivative terms, Hamiltonian (\ref{NNNAFM}) becomes $H=\int dx \,{\mathcal H}$, where 
\bea
\label{hmn}
{\mathcal H} = \sum_{i=1}^N {\mathcal H}_i^{\rm 1D} & + &  J_{\perp}a_0\left({\bf M}_i\cdot{\bf M}_{i+1} + {\bf N}_i\cdot{\bf N}_{i+1}\right) \nonumber \\
& + & 2J_{\times}a_0\left({\bf M}_i\cdot{\bf M}_{i+1}- {\bf N}_i\cdot{\bf N}_{i+1}\right). 
\eea
In terms of the currents, ${\bf M}_i\equiv{\bf J}_i+{\bar {\bf J}}_i$,  the Hamiltonian density (\ref{hmn}) may be written:\cite{Allen:Fate,Nersesyan:Spinons}
\bea
{\mathcal H}  =  \sum_{i=1}^N  {\mathcal H}_i^{\rm 1D} & + & \lambda_1({\bf J}+\bar {\bf J})_i\cdot({\bf J}+\bar {\bf J})_{i+1}\nonumber \\
& +& \lambda_2{\bf N}_i\cdot{\bf N}_{i+1}+{\cdots}
\eea
where
\be
\lambda_1=(J_{\perp}+2J_{\times})a_0, \quad \lambda_2=(J_{\perp}-2J_{\times})a_0.\ee
In particular, for $J_\times=J_\perp/2$, the strongly relevant interchain coupling, $\lambda_2$, between the staggered magnetizations vanishes.\cite{Allen:Fate,Nersesyan:Spinons} Setting $J_\times=J_\perp/2$, and neglecting velocity renormalizing terms, the Hamiltonian splits into two independent pieces, or ``parity'' sectors:\cite{Allen:Fate,Nersesyan:Spinons} 
\be
{\mathcal H}={\mathcal H}_+ +{\mathcal H}_-,
\ee
where
\bea
{\mathcal H}_+& =  & \sum_{i} {\mathcal N}\,\hbar v\,({\bf J}_{2i}\cdot{\bf J}_{2i}+{\bf \bar J}_{2i+1}\cdot{\bf \bar J}_{2i+1}) \nonumber \\
& & \hspace{2cm} + \lambda_1 {\bf J}_{2i}\cdot{\bf \bar J}_{2i+1},
\eea
and ${\mathcal H}_-$ is obtained from ${\mathcal H}_+$ by the (parity) transformation ${\bf J}\leftrightarrow{\bf \bar J}$. In the sector of positive parity, the even (odd) chains carry left (right) moving fields; in the sector of negative parity the situation is reversed (see figure \ref{fig:plusparity}). Equivalently, ${\mathcal H}_+$ and  ${\mathcal H}_-$ are interchanged under a shift by $a_0$ transverse to the chains.
Specializing to the ladder:
\be
\label{twohplus}
{\mathcal H}_+   =  {\mathcal N}\hbar v\,({\bf \bar J}_{\rm I}\cdot{\bf \bar J}_{\rm I}+ {\bf J}_{\rm II}\cdot{\bf J}_{\rm II})+\lambda_1({\bf \bar J}_{\rm I}\cdot{\bf J}_{\rm II}), 
\ee
where we label the chains by Roman numerals to avoid subsequent confusion with space-time indices.
The Hamiltonian (\ref{twohplus}) may be brought into a more familiar form by introducing a spinor, the left component of which resides on one chain and the right resides on the other:
\be
\label{twochainspinor}
\psi_+=\begin{pmatrix} {{\mathcal R}_{\rm I} \cr  {\mathcal L}_{\rm II}}\end{pmatrix}.
\ee
In terms of this spinor, the Hamiltonian (\ref{twohplus}) becomes
\be
\label{posthirr}
{\mathcal H}_+   = {\mathcal N}\hbar v\,({\bf J}_+\cdot{\bf J}_+ + {\bf \bar J}_+\cdot{\bf \bar J}_+ )+\lambda_1({\bf J}_+\cdot{\bf \bar J}_+)
\ee
and similarly for ${\mathcal H}_-$. (Equivalently one may perform the {\it chiral} interchange ${\bf J}_{\rm I} \leftrightarrow {\bf J}_{\rm II}$ on the original Hamiltonians.) We see that ${\mathcal H}_+$ is nothing but an \new{${\rm SU}(\rm 2)$} Thirring model. That is to say, {\it the frustrated ladder may be reformulated as the sum of two decoupled \new{${\rm SU(\rm 2)}$} Thirring models}, labelled by their parity. \cite{Allen:Fate,Nersesyan:Spinons} We emphasize that each of these decoupled models captures the behavior of the coupled ladder, as highlighted in (\ref{twochainspinor}), and not just a single chain. In particular, the elementary excitations of the ladder are those of the \new{$\rm{SU(2)}$} Thirring model, namely massive spinons. These correspond to domain walls separating regions of different spontaneous dimerization.\cite{Allen:Fate}

\new{In this paper we straightfowardly replace the ${\rm SU(2)}$ currents by ${\rm SU(N)}$ currents, as suggested in reference \onlinecite{Allen:Fate}. In each parity sector, the Hamiltonian becomes that of the ${\rm SU(N)}$ Thirring model with ${\rm N}-1$ multiplets of massive spinons (see appendix \ref{NATM}). This is the simplest generalization which retains spinon excitations and parity sectors. We note that the alternative strategy of replacing lattice spins by ${\rm SU(N)}$ generators leads to problems at the outset.\cite{Lecheminant:Private} As we discuss in appendix \ref{app:spin}, the representation of the generators translates into the filling of the corresponding electronic model. For the critical ${\rm SU(N)}$ Heisenberg model, with spins in the lowest fundamental representation, the corresponding Hubbard model has one electron per site.\cite{Assaraf:MIT} The corresponding ``spin'' density (\ref{magstag}) has harmonics at multiples of $2k_F=2\pi/{\rm N}a_0$ due to all the fundamental primaries of the $\widehat{\rm su}(\rm N)_1$ WZNW model. In this case, the simple fine tuning condition, $J_\times=J_\perp/2$, does not remove all relevant perturbations.\cite{Lecheminant:Private}
The absence of such terms is crucial for spinons in the confederate flag model, and such a generalization would be inappropriate. Attempts to reinstate the condition of half-filling with Hubbard chains or the alternating ${\rm N}\otimes\bar {\rm N}$ magnet (${\rm q}={\rm N}^2$ quantum Potts model) also lead to difficulties; for ${\rm N}>2$ they are massive and dimerized\cite{Affleck:LargeN,Marston:LargeN,Affleck:Dimerization,Read:Some} and have little in common with the UV limit of decoupled spin-$1/2$ chains. Since our interest in these generalized models stems from the spinon physics of the confederate flag model, we confine ourselves to the simple minded extension of the low energy action. We study the ${\rm SU(N)}$ evolution of the original operators, and retain the terms smooth and staggered magnetizations for these fields.}

In the next section, we shall compute the dynamical structure factors of the generalized \new{model}. These are a direct probe of the elementary excitations.
\section{Dynamical Structure Factor}
\label{sect:dsf}
In this section we compute the dynamical structure factor (as may be seen by neutrons) for momentum transfers close to the ``soft modes'' at $0$ and ${\pi}$. This is nothing but a Fourier transform of the spin-spin correlation functions:
\bea
\label{dsfdef}
S(\omega,q,q_\perp) & \propto & {\rm Im} \, i\int_{-\infty}^{\infty} dx\int_0^\infty dt \, e^{i(\omega+i\delta) t-ivqx}  \nonumber \\ 
& & \hspace{-2cm} \langle\,[S_{\rm I}^a(t,x)\pm S_{\rm II}^a(t,x),S_{\rm I}^a(0,0)\pm S_{\rm II}^a(0,0)]\,\rangle.
\eea 
The plus (minus) sign corresponds to $q_\perp=0$ ($q_\perp=\pi$), and $\delta$ ensures convergence of the temporal integral. The longitudinal momentum transfers in the vicinity of $q=0$ ($q=\pi$) probe the smooth (staggered) components of the spin operators. The task is to relate the spin operators entering (\ref{dsfdef}) to the operators of the Thirring models, and to evaluate their matrix elements. 

\subsection{Smooth Components}
\begin{figure}
\begin{center}
\psfrag{s2}{$s^2/m^2$}
\psfrag{SF}[ct][ct]{$S(\omega,q\sim 0,0)$}
\noindent
\epsfysize=0.25\textwidth
\epsfbox{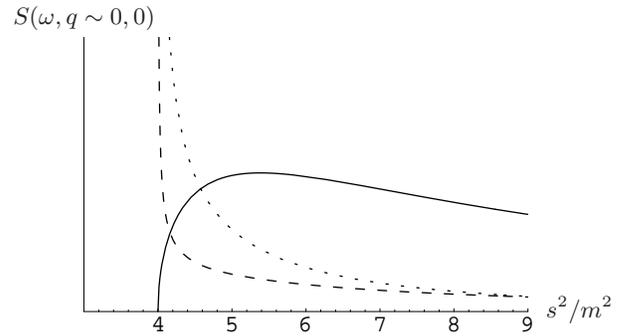}
\end{center}
\caption{\label{fig:sfcompare} Exact dynamical structure factor $S(\omega,q\sim 0,0)$ for ${\rm N}=2$ (solid), ${\rm N}=3$ (dashed) and ${\rm N}=\infty$ (dotted) with fixed $q$ and arbitrary normalization. The $\sqrt{s^2-4m^2}$ threshold behavior for the ${\rm N}=2$ physical case, is replaced by a  $1/\sqrt{s^2-4m^2}$ divergence for all ${\rm N}>2$.}
\end{figure}

The smooth component of the sum of the chain spin densities may be expressed in terms of the two Thirring models as follows:
\bea
{\bf S}_{\rm I}+{\bf S}_{\rm II}|_{\rm smooth} & = & {\bf J}_{\rm I}+{\bf\bar J}_{\rm I}+{\bf J}_{\rm II}+{\bf\bar J}_{\rm II} \nonumber \\
& \equiv & {\bf J}_{0,+}+{\bf J}_{0,-} \nonumber 
\eea
where ${\bf J}_{0,+}={\bf \bar J}_{\rm I}+{\bf J}_{\rm II}$ (${\bf J}_{0,-}={\bf J}_{\rm I}+{\bf \bar J}_{\rm II}$) is the temporal component of the Thirring current in the model of positive (negative) parity. Simply put, the structure factor $S(w,q\sim0,0)$ of the frustrated ladder, may be obtained from the correlators of ${\bf J}_0$ in the ${\rm SU(N)}$ Thirring model.
\bea
\label{s00j}
S(\omega,q\sim0,0) & \propto & {\rm Im} \sum_{{\mathcal P}=\pm}\,i\int_{-\infty}^{\infty} dx\int_0^\infty dt \,e^{i(\omega+i\delta)t-ivqx}  \nonumber \\ 
& & \langle\,[J_{0,{\mathcal P}}^a(t,x),J_{0,{\mathcal P}}^a(0,0)]\,\rangle
\eea
where the summation is over parity sectors. The elementary excitations of the ${\rm SU(N)}$ Thirring model consist of ${\rm N}-1$ multiplets of massive particles, corresponding to the fundamental representations of ${\rm SU(N)}$. The length of the Young tableau is termed the ``rank'' of the particle,\cite{Smirnov:Form} and their masses are given by (\ref{massfun}). It is convenient to move to a basis of such particles and to parametrize their energy and momentum in terms of rapidity:
\be
E_i=m_i\cosh\theta_i \quad P_i=m_i\sinh\theta_i.
\ee
One may now insert a complete set of states between the current operators in (\ref{s00j})\be
{\mathbb I}=\sum_{n=0}^\infty\sum_{\epsilon_i}\int\frac{d\theta_1\dots d\theta_n}{(2\pi)^2n!}\,|\theta_n\dots\theta_1\rangle_{\epsilon_n\dots\epsilon_1}\ ^{\epsilon_1\dots\epsilon_n} \langle\theta_1\dots\theta_n|
\ee
where the ${\epsilon}_i$ are the internal (or isotopic) indices carried by the members of each multiplet. Using 
\bea
& & \hspace{-1cm} \ ^{\epsilon_1^\prime\dots\epsilon_n^\prime}\langle\theta_1^\prime\dots\theta_n^\prime|{\mathcal O}(t,x)|\theta_n\dots\theta_1\rangle_{\epsilon_n\dots\epsilon_1} \nonumber \\
& & \hspace{0.5cm}\equiv e^{i\sum_j(E_j^\prime-E_j)t-(P_j^\prime-P_j)x}\times \nonumber \\
& &  \hspace{1cm}\ ^{\epsilon_1^\prime\dots\epsilon_n^\prime}\langle\theta_1^\prime\dots\theta_n^\prime|{\mathcal O}(0,0)|\theta_n\dots\theta_1\rangle_{\epsilon_n\dots\epsilon_1}
\eea
one obtains
\bea
\label{intxt}
S(\omega,q\sim0,0) & \propto & \nonumber \\
  & & \hspace{-3.2cm}-2\pi \,{\rm Im} \sum_{n=0}^\infty\sum_{\epsilon_i}\int\frac{d\theta_1\dots d\theta_n}{(2\pi)^n n!}\,|F_{J_0^a}(\theta_1\dots\theta_n)_{\epsilon_1\dots\epsilon_n}|^2  \nonumber \\
& & \hspace{-3.2cm}\left[\frac{\delta(vq-\sum_jm_j\sinh\theta_j)}{\omega-\sum_j m_j\cosh\theta_j+i\delta}-\frac{\delta(vq+\sum_j m_j\sinh\theta_j)}{\omega+\sum_j m_j\cosh\theta_j+i\delta}\right]
\eea
where $F_{J_0^a}(\theta_1\dots\theta_n)_{\epsilon_1\dots\epsilon_n}$ is a multiparticle form factor of the temporal Thirring current:
\be
F_{J_0^a}(\theta_1\dots\theta_n)_{\epsilon_1\dots\epsilon_n}\equiv\langle 0|J_0^a(0,0)|\theta_n\dots\theta_1\rangle_{\epsilon_n\dots\epsilon_1}.
\ee
The dominant contributions to (\ref{intxt}) come from the states with the lowest mass. In the case at hand these are two particle states of the (rank-1) fundamental ${\Box}$, and it's (rank-N-1) conjugate ${\bar\Box}$. In particular, the current operator couples to the adjoint representation occuring in the ${\rm SU(N)}$ tensor product  $\Box\otimes\bar\Box$; for ${\rm N}=2$, ${\bar\Box}$ is ${\Box}$. 
As we discuss in appendix \ref{NATM}, this form factor is
\be
F_{J_0}(\theta_1,\theta_2)_{\Box,\bar\Box}\propto m\sinh\left(\frac{\theta_1+\theta_2}{2}\right)f^{\Box\bar\Box}_{\rm adj}(\theta_{12})
\ee
where
\bea
\label{fofn}
f^{\Box\bar\Box}_{\rm adj}(\theta_{12}) & = & \nonumber \\ 
& & \hspace{-2.6cm}\exp\left\{\int_0^\infty dx\,\frac{2\exp(x/{\rm N})\sinh(x/{\rm N})\sin^2(x\hat\theta/2\pi)}{x\sinh^2x}\right\}
\eea
and $\hat\theta=i\pi-\theta$; see equations (\ref{temporal}) and  (\ref{fminint}). We have supressed the isotopic and component information in (\ref{fofn}) and concentrated solely on the rapidity dependence. Inserting this into (\ref{intxt}) and performing the $\theta$ integrations one obtains
\be
\label{szerozero}
S(\omega,q\sim 0,0)\propto\frac{m^2 v^2 q^2}{s^3\sqrt{s^2-4m^2}}\,|f^{\Box\bar\Box}_{\rm adj}[2\theta(s)]|^2
\ee
where $s^2=\omega^2-v^2q^2$, $\theta(s)={\rm arcosh}(s/2m)$ and 
\be
\label{bsthresh}
4m^2<s^2<\left\{\begin{array}{ll}
16m^2 & \mbox{${\rm N}=2$}, \\
9m^2 & \mbox{${\rm N}=3$}, \\
16m^2\cos^2(\pi/{\rm N}) & \mbox{${\rm N}>3$}. 
\end{array} \right.
\ee 
This result is plotted in figure \ref{fig:sfcompare} and is exact, provided (\ref{bsthresh}) is fulfilled. For larger energy transfers there are small corrections due to higher mass states; the upper thresholds correspond to four rank-1 solitons, three rank-1 (or rank-2) solitons, and a rank-2 bound state and its conjugate, respectively.
In particular, there are no single-particle bound states appearing {\it below} the gap; the elementary Thirring excitations correspond to fundamental ${\rm SU(N)}$ representations, and do not couple to the current directly, which spans the adjoint.

The result (\ref{szerozero}) interpolates between two known limits. For ${\rm N}=2$, it coincides with equation (34) of reference \onlinecite{Allen:Fate}, and in the limit ${\rm N}=\infty$, where (\ref{fofn}) tends to unity, we recover the result for free massive fermions.\cite{Allen:Frustrated,Allen:Fate}  In particular, the ($\theta=0$) threshold behavior of (\ref{fofn}) is quite instructive: for ${\rm N}=2$ it vanishes like $\sinh(\theta/2)$, as may be seen from (\ref{fabform}), whereas it is finite and non-vanishing for {\it any} ${\rm N}>2$. As a result, the structure factor (\ref{szerozero}) vanishes as $\sqrt{s^2-4m^2}$ in the physical case of ${\rm N}=2$, but diverges as $1/\sqrt{s^2-4m^2}$ for any ${\rm N>2}$ --- see figure \ref{fig:sfcompare}. Solely on the basis of the ${\rm N}=2$ and ${\rm N}=\infty$ limits,\cite{Allen:Fate,Allen:Frustrated} one might have expected the threshold to get steeper and narrower with increasing ${\rm N}$, but to remain qualitatively correct for ${\rm N}<\infty$. The actual evolution, and the departure even for ${\rm N}=3$, is a sobering example of how ${\rm SU(N)}$ treatments may miss simple features over the entire range of ${\rm N}$.

Likewise, the smooth component of the difference of the chain spin densities may be expressed in terms of the two Thirring models as follows:
\bea
{\bf S}_{\rm I}-{\bf S}_{\rm II}|_{\rm smooth} & = & {\bf J}_{\rm I}+{\bf\bar J}_{\rm I}-{\bf J}_{\rm II}-{\bf\bar J}_{\rm II} \nonumber \\
& \equiv & {\bf J}_{1,+}-{\bf J}_{1,-} \nonumber 
\eea
where ${\bf J}_{1,+}={\bf \bar J}_{\rm I}-{\bf J}_{\rm II}$ (${\bf J}_{1,-}={\bf J}_{\rm I}-{\bf \bar J}_{\rm II}$) is the spatial component of the Thirring current in the model of positive (negative) parity. Simply put, the structure factor $S(w,q\sim0,\pi)$ of the frustrated ladder, may be obtained from the correlators of ${\bf J}_1$ in the ${\rm SU(N)}$ Thirring model. The corresponding form factor is given by (\ref{spatial}):
\be
F_{J_1}(\theta_1,\theta_2)_{\Box,{\bar\Box}}\propto m\cosh\left(\frac{\theta_1+\theta_2}{2}\right)f^{\Box\bar\Box}_{\rm adj}(\theta_{12}).
\ee
We obtain
\be
S(\omega,q\sim 0,\pi)\propto\frac{m^2 \omega^2}{s^3\sqrt{s^2-4m^2}}\,|f^{\Box\bar\Box}_{\rm adj}[2\theta(s)]|^2.
\ee
Once again, this result interpolates between the known ${\rm N}=2$ and ${\rm N}=\infty$ results,\cite{Allen:Fate,Allen:Frustrated} and \new{the ${\rm SU(N)}$ approach} leads to qualitatively incorrect results over the entire range of ${\rm N}>2$.

\subsection{Staggered Components}
We denote the staggered component of the spin on chain ${\rm I}$, ${\bf S}_{\rm I}(t,x)|_{\rm stagg.}$ by ${\bf N}_{\rm I}(t,x)$. In the UV limit (corresponding to decoupled chains and $m=0$) ${\bf N}(t,x)$ is a spinless  ${\widehat {\rm su}}({\rm N})_1$ primary field with (full) scaling dimension $\Delta=1-1/{\rm N}$. For the ladder we propose the following formula for the long distance asymptotics of the real space correlation functions:
\bea
\label{staggbessel}
& &  \hspace{-0.7cm}\langle [{\bf N}_{\rm I}(t,x) \pm  {\bf N}_{\rm II}(t,x)].[{\bf N}_{\rm I}(0,0)\pm {\bf N}_{\rm II}(0,0)]\rangle \nonumber \\
 & & \hspace{0.2cm}\propto \langle {\bf N}_{\rm I}(t,x).{\bf N}_{\rm I}(0,0)\rangle\pm \langle {\bf N}_{\rm I}(t,x).{\bf N}_{\rm II}(0,0)\rangle \nonumber \\
 & & \hspace{0.2cm} \propto  m^{2\Delta}\,[K_{\Delta}^2(mr)\pm K_0^2(mr)] +\cdots
\eea
where $r\equiv\sqrt{z\bar z}=\sqrt{x^2-t^2}$ ($v=1$) and $K_\nu(x)$ is Macdonald's function,\cite{Lebedev:Special} also known as the modified Bessel function of the third kind.\cite{Bateman2} The dots stand for more rapidly decaying terms.
In order to get a feel for this result we begin by studying a few limits.
In the limit ${\rm N}\rightarrow \infty$, $\Delta\rightarrow1$, each parity sector reduces to non-interacting massive fermions. More specifically, ${\bf N}_i$ may be replaced by the fermion bilinear ${{\mathcal L}^\dagger_{i,\alpha}{\bf t}_{\alpha\beta} {\mathcal R}_{i,\beta}+{\mathcal R}^\dagger_{i,\alpha} {\bf t}_{\alpha\beta}{\mathcal L}_{i,\beta}}$ and one obtains
\bea
\label{interferm}
\langle{\bf N}_{\rm I}.{\bf N}_{\rm I}\rangle & \propto & \langle {\mathcal L}^\dagger_{\rm I} {\mathcal L}_{\rm I}\rangle\langle {\mathcal R}_{\rm I}{\mathcal R}_{\rm I}^\dagger\,\rangle \\ 
\label{intraferm}
\langle{\bf N}_{\rm I}.{\bf N}_{\rm II}\rangle & \propto & \langle {\mathcal L}^\dagger_{\rm I} {\mathcal R}_{\rm II}\rangle\langle {\mathcal R}_{\rm I}{\mathcal L}_{\rm II}^\dagger\,\rangle
\eea
with the usual massive Dirac fermion correlators:
\bea
\langle {\mathcal L}^\dagger {\mathcal L}\rangle & = & 2m\sqrt{\frac{\bar z}{z}}K_1(mr) \\
\langle {\mathcal L}^\dagger {\mathcal R}\rangle & = & 2mK_0(mr)
\eea
--- see for example chapter 13 of the book \onlinecite{Tsvelik:QFT}. In equations (\ref{interferm}) and (\ref{intraferm}) we see quite clearly that the correlators of staggered magnetizations are {\it products} of correlators from the sectors of {\it different} parity;\cite{Allen:Fate,Nersesyan:Spinons} by definition the left and right moving fields on a given chain belong to different sectors. In coupling to the staggered magnetizations, the solitons are still created in pairs, but belong to different sectors. \cite{Allen:Fate,Nersesyan:Spinons} In a given sector (i.e. Thirring model) we thus require the matrix elements of {\it single-soliton} creation operators. The matrix elements of such operators have only recently become available.\cite{Lukyanov:Soliton,Essler:Weakly,Essler:Spectral} The free fermions appearing in (\ref{interferm}) and (\ref{intraferm}) for ${\rm N}\rightarrow\infty$, are replaced by chiral fields $L_s,R_s$, which are non-local single soliton creation operators and carry the Lorentz spin, $\pm\Delta/2$, of a Thirring soliton;\cite{Halpern:Quantum,Koberle:Scattering} we take the plus (minus) sign for left (right) movers. These chiral fields are the components of an (interacting) $\widehat{{\rm su}}({\rm N})_1$ primary field, and the Lorentz spin is nothing but the UV conformal dimension. The single-solition form factors of such operators are governed (upto normalization) solely by their Lorentz transformation properties:
\be
\langle 0 |{\mathcal L}_s|\theta\rangle=m^{\Delta/2}e^{\Delta\theta/2},\quad \langle 0 |{\mathcal R}_s|\theta\rangle=m^{\Delta/2}e^{-\Delta\theta/2},
\ee
and their two-point functions are now readily computed:
\bea
\label{intkdelta}
\langle {\mathcal L}_s^\dagger {\mathcal L}_s\rangle & = & m^\Delta\int d\theta \, e^{\Delta\theta}\, e^{-\tau m{\rm ch}\theta+ixm{\rm sh}\theta} \\
\label{sunll}
& = & m^{\Delta}\left(\frac{\bar z}{z}\right)^{\Delta/2}\,2K_\Delta(m\sqrt{z\bar z})\\
\label{intkzero}
\langle {\mathcal L}^\dagger_s {\mathcal R}_s\rangle & = & m^\Delta \int d\theta \,e^{-\tau m{\rm ch}\theta+ixm{\rm sh}\theta}\\
\label{sunlr}
& = & m^\Delta\, 2K_0(m\sqrt{z\bar z})
\eea
where $z=\tau-ix$ and $\tau=it$. The results for $\langle {\mathcal R}^\dagger_s{\mathcal R}_s\rangle$ and $\langle {\mathcal R}^\dagger_s {\mathcal L}_s \rangle$ follow by interchanging $z$ and $\bar z$.
In particular, equation (\ref{sunll}) first appeared  in the study of weakly coupled one-dimensional Mott insulators.\cite{Essler:Weakly} Replacing the correlators in (\ref{interferm}) and (\ref{intraferm}) with these more general expressions, the result (\ref{staggbessel}) follows immediately. 

Further, the Macdonald function has the asymptotic expansion given by equation 9.7.2 of reference \onlinecite{Abramowitz:Tables}:
\bea
\label{mcexpansion}
& & \hspace{-0.5cm} K_{\Delta}(mr)  =   \nonumber  \\ 
& &  \hspace{-0.3cm} \sqrt{\frac{\pi}{2mr}}e^{-mr}\left[1+\frac{\mu-1}{8mr}+\frac{(\mu-1)(\mu-9)}{2!\,(8mr)^2}+\dots\right]
\eea
where $\mu=4\Delta^2$. The leading term in (\ref{mcexpansion}) is independent of $\Delta$, and at separations $r\gg 1/m$, the interchain and intrachain correlations (amusingly) coincide: 
\be
\langle{\bf N}_a(t,x).{\bf N}_b(0,0)\rangle\sim \frac{m^{2\Delta-1}}{r}e^{-2mr}.\ee
Coupling the chains together not only generates exponentially decaying interchain correlations, but also modifies the $1/r^{2\Delta}$ behavior within the chains.

\begin{figure}
\begin{center}
\psfrag{s2}{$s^2/m^2$}
\psfrag{SF}[ct][ct]{$S(\omega,q\sim \pi,0)$}
\noindent
\epsfysize=0.25\textwidth
\epsfbox{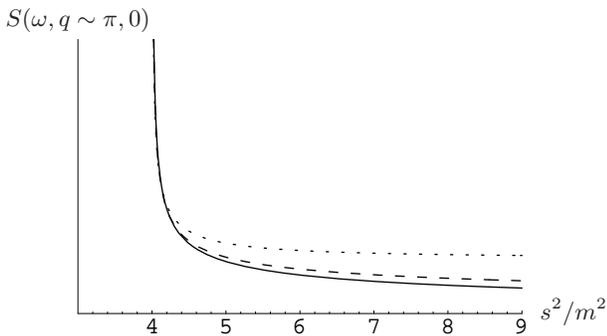}
\end{center}
\caption{\label{fig:sfcompare2} Exact dynamical structure factor $S(\omega,q\sim \pi,0)$ for ${\rm N}=2$ (solid), ${\rm N}=3$ (dashed) and ${\rm N}=\infty$ (dotted). The threshold behavior is $1/\sqrt{s^2-m^2}$ for all ${\rm N}$, and we have normalized accordingly. The ${\rm SU(N)}$ approach leads to qualitatively similar results over the entire range of ${\rm N}$.}
\end{figure}

Substituting (\ref{staggbessel}) into the definition (\ref{dsfdef}) and effecting the Fourier transforms we obtain the following structure factors:
\be
\label{spizero}
S(\omega,q\sim \pi,0)\propto\frac{\left[s+\sqrt{s^2-4m^2}\right]^{2\Delta}+(2m)^{2\Delta}}{s\sqrt{s^2-4m^2}}
\ee
\be
\label{spipi}
S(\omega,q\sim \pi,\pi)\propto\frac{\left[s+\sqrt{s^2-4m^2}\right]^{2\Delta}-(2m)^{2\Delta}}{s\sqrt{s^2-4m^2}}
\ee
where $s^2=\omega^2-(q-\pi)^2$. In deriving these expressions the reader may find the integral representations (\ref{intkdelta}) and (\ref{intkzero}) more convenient. At threshold, $S(\omega,q\sim \pi,0)$ diverges as $1/\sqrt{s^2-4m^2}$ for {\it all} ${\rm N}$, and we plot this behavior in figure \ref{fig:sfcompare2}; the large $s$ behavior is $s^{-2/{\rm N}}$. Similarly, at threshold, $S(\omega,q\sim \pi,\pi)$ tends to a constant for {\it all} ${\rm N}$. In contrast to the magnetization correlators, we obtain qualitatively similar results over the entire range of ${\rm N}$.

\section{Conclusions}
\label{sect:conc}
In this paper we have studied the ${\rm SU(N)}$ evolution of a frustrated spin ladder.\cite{Allen:Fate,Nersesyan:Spinons}  The dynamical structure factors corresponding to the smooth (staggered) magnetizations are shown to be qualitatively different (the same) in the ${\rm N}=2$ and ${\rm N}>2$ cases. A robust feature which survives however, is the absence of coherent single-particle excitations at low energies. This is in stark contrast to the unfrustrated ladder, and reinforces the notion of spinons stabilized by frustration.\cite{Allen:Fate,Nersesyan:Spinons}

In closing we note that the ${\rm N}=\infty$ limit of the two chain model is a free theory, but this is not so in general. In particular, each parity sector of the four chain model may be viewed as two (non-chiral) $\widehat{\rm su}(\rm N)_1$ WZNW models coupled by currents:
\bea
{\mathcal H}_+ &=& \widehat{\rm su}(\rm N)_1+\widehat{\rm su}(\rm N)_1+\lambda J_{+}{\bar J}_{+}.\nonumber   
\eea
With transverse periodic boundary conditions the currents
${\bf \bar J}_+={\bf \bar J}_{\rm I}+{\bf \bar J}_{\rm III}$, ${\bf J}_+={\bf J}_{\rm II}+{\bf J}_{\rm IV}$ generate $\widehat{\rm su}(\rm N)_2$ Kac--Moody algebras, and it is convenient to write:\cite{Nersesyan:Spinons,ZF:su2,ZF:Nonlocal}
\bea
{\mathcal H}_+ & = & {\mathbb Z}_{\rm N}+ (\widehat{\rm su}(\rm N)_2+\lambda J_{+}{\bar J}_{+}).\nonumber
\eea
The ${\mathbb Z}_{\rm N}$ parafermions\cite{ZF:Nonlocal} describe gapless non-magnetic excitations with central charge ${\rm c}=2({\rm N}-1)/({\rm N}+2)$. They are unaffected by the interaction (due to the boundary conditions) and for ${\rm N}\rightarrow\infty$ they reduce to two Gaussian models. The $\widehat{\rm su}(\rm N)_2$ model on the other hand is rendered massive by the interaction, and is in fact integrable. The mass spectrum coincides with that of the (two chain) ${\rm SU(N)}$ Thirring model (\ref{massfun}), but the scattering is notably different. The S-matrix can be extracted from a straightforward generalization of the Thermodynamic Bethe Ansatz (TBA) equations derived in reference \onlinecite{Tsvelik:SUSY}. At low temperatures ($T\ll M_1$) the free energy of the perturbed $\widehat{\rm su}(\rm N)_k$ WZNW model is given by
\bea
F/L=-T\sum_{j=1}^{{\rm N}-1}M_j\int\frac{d\theta}{2\pi}\,{\rm ch}\theta\,\ln[1+e^{-\varepsilon_k^{(j)}(\theta)/T}]\nonumber
\eea
where in this case $k=2$. The excitation energies $\varepsilon_n^{(j)}$ ($j=1,\dots {\rm N}-1$, $n=1,2\dots$) satisfy
\bea
\hspace{-1cm}&&T\ln(1+e^{\varepsilon_n^{(i)}(\lambda)/T}) \nonumber \\
&&\hspace{0.5cm}-T{\mathcal A}_{ij}\ast C_{nm}\ast\ln(1+e^{-\varepsilon_m^{(j)}(\lambda)/T})\nonumber \\
&&\hspace{3cm}=\delta_{n,k}M_i\,{\rm ch}(2\pi\lambda/N)\nonumber
\eea
where $\ast$ denotes convolution,
the kernels $C_{nm}(\lambda)$ and ${\mathcal A}_{ij}(\lambda)$ are given in reference \onlinecite{Tsvelik:SUSY} and $\lambda={\rm N}\theta/2\pi$.
We extract the Bethe equations $E=\sum_{a=1}^n m\,{\rm ch}\,\theta_a$ and
\bea
&&\exp(imL\,{\rm sh}\,\theta_a)=\nonumber \\
&&\hspace{0.5cm}\prod_{b\neq a} S_0(\theta_{a}-\theta_b)\prod_{\alpha}e_1(\theta_a-\lambda_\alpha)\prod_\beta{\mathcal E}(\theta_a-\mu_\beta)\nonumber
\eea
where
\bea
&& S_0(\theta)=\exp\left\{-\int_{-\infty}^\infty \frac{d\omega}{\omega}e^{-i\theta\omega}\times \right. \nonumber \\
&&\left.\quad \left[-1+\frac{1}{(1+e^{-2|\omega|\pi/{\rm N}})^2}\left(1+\frac{{\rm sh}\,\pi(1-\frac{2}{{\rm N}})\omega}{{\rm sh}\,\pi\omega}\right)\right]\right\},\nonumber
\eea
and 
\bea
e_n(x)=\frac{x-i\pi n/{\rm N}}{x+i\pi n/{\rm N}},\quad {\mathcal E}(x)=\frac{e^{{\rm N}x/2}-i}{e^{{\rm N}x/2}+i}.\nonumber 
\eea
The rapdities $\lambda_{\alpha}$ and $\mu_{\beta}$ are distributed according to the $A^{{\rm N}-1}$ heirarchy, the details of which do not concern us here. Similar equations occur for the ${\rm SU(N)}$ invariant Thirring model ($k=1$) but without the $\mu$ rapidities and with a different $S_0(\theta)$. In the limit ${\rm N}\rightarrow\infty$ one obtains
\bea
\exp(imL\,{\rm sh}\,\theta_a)=\prod_{b\neq a} S_0^\infty(\theta_{ab})\nonumber
\eea
where $S_0^{\infty}(\theta)=\exp\left(-i\frac{\pi}{2}\,{\rm sign}\,\theta\right)$. This is to be contrasted with the ${\rm SU(N)}$ invariant Thirring model where $S_0^\infty(\theta)=-1$. The absence of a simple ${\rm N}\rightarrow\infty$ limit will be crucial for multiparticle form factors, and renders excitations with non-trivial statistics. In future publications we hope to study these pertinent issues in more detail. Recent progress on spinon propagation in the four chain model may be found in the work of Smirnov and Tsvelik.\cite{Smirnov:Propagating}
\begin{center}
{\bf Acknowledgements}
\end{center}
We are extremely grateful to Fabian Essler and Feodor Smirnov. \new{We are indebted to Philippe Lecheminant for valuable comments.} We thank Sam Carr for proof reading this manuscript. We also acknowledge support from the US DOE under contract number DE-AC02 -98 CH 10886. 
\appendix
\section{Spin Operators}
\label{app:spin}
\new{In this appendix we comment on the connection between ${\rm SU(N)}$ spin representations and filling.} At each lattice site (labelled by $n$) one may introduce the fermionic spin operators
\be
\label{spinfermi}
{S}_n^a=\sum_{\alpha,\beta=1}^{\rm N}c_{n,\alpha}^\dagger t_{\alpha\beta}^a c_{n,\beta}
\ee
where $c$ and $c^\dagger$ obey the canonical fermionic anticommutation relations\be
\label{acr}
\{c_{n,\alpha}^\dagger,c_{m,\beta}\}=\delta_{n,m}\delta_{\alpha,\beta}\quad \{c,c\}=0\quad\{c^\dagger,c^\dagger\}=0
\ee
and the generators $t^a$ span the algebra ${\rm su}(\rm N)$: $[t^a,t^b]=i{f^{ab}}_c\,t^c$. It is readily verified that spins on different sites commute, whereas those on the same site satisfy the ${\rm su}(\rm N)$ algebra: $[S_n^a,S_m^b]=i\delta_{n,m}{f^{ab}}_c\,S_n^c$. In the fundamental representation, the generators are chosen to satisfy
$
{\rm tr}(t^at^b)={\mathcal C}\,\delta^{ab}$, $t^at^a={\mathcal C}_2\,{\mathbb I},
$
with ${\mathcal C}=1/2$ and ${\mathcal C}_2=({\rm N}^2-1)/2{\rm N}$; see appendix A.3 of \onlinecite{Peskin}. 

\begin{figure}[ht]
\begin{center}
\noindent
\epsfysize=0.1\textwidth
\epsfbox{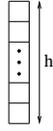}
\end{center}
\caption{\label{fig:young} 
Antisymmetric ${\rm su}(\rm N)$ Young tableau of height $h$.
} 
\end{figure}

One may specify the ${\rm su}(\rm N)$ representation on which spin operators ${\bf S}_n$ act by the relevant Young tableau --- see for example \onlinecite{Jones:Groups}. In particular, this fixes the value of the quadratic Casimir ${\bf S}_n^2$, and thus by equation (\ref{spinfermi}), constrains fermion occupation numbers. As we shall demonstrate, the constraint
\be
\label{constraint}
\sum_{\alpha=1}^{\rm N}c_{n,\alpha}^\dagger c_{n,\alpha}=h; \quad \forall n.
\ee
corresponds to the vertical (i.e. antisymmetric) Young tableau of height $h$, as depicted in figure \ref{fig:young}.
The constraint (\ref{constraint}) fixes $h$ electrons per site, and the permissable states to be of the form:
\be
\label{states}
\psi_{\alpha_1,\alpha_2,\cdots,\alpha_h}=c^\dagger_{n,\alpha_1}c^\dagger_{n,\alpha_2}\cdots c_{n,\alpha_h}^\dagger|0\rangle,
\ee
where $\alpha_i\in [1,\cdots,{\rm N}]$. By virtue of the fermion anticommutation relations (\ref{acr}), this may be viewed as a tensor of rank $h$, antisymmetric under the interchange of any pair of labels $\alpha$; by the standard conventions for Young tableau \cite{Jones:Groups} this corresponds to a vertical diagram of $h$ boxes. Moreover, it also follows from the anticommutation relations (\ref{acr}), that there are ${\rm N}({\rm N}-1)\cdots({\rm N}-h+1)/h!$ independent states of the form (\ref{states}); this coincides with the dimension of the representation corresponding to the Young tableau of fig.\ref{fig:young}; see \S 8.4 of \onlinecite{Jones:Groups}. Further, squaring equation (\ref{spinfermi}) and enforcing the constraint (\ref{constraint}) one obtains\footnote{Note the identity $t^a_{ij}t^a_{kl}=\frac{1}{2}(\delta_{il}\delta_{kj}-\delta_{ij}\delta_{kl}/{\rm N})$.}
\be
{\bf S}_n^2=\frac{h({\rm N}^2-h)}{2{\rm N}}+\frac{h(1-h)}{2}.
\ee
This coincides with the quadratic Casimir of the ${\rm su}({\rm N})$ Young tableau depicted in fig.\ref{fig:young};\cite{Okubo:Casimir} see eq. 2.19 of \onlinecite{Andres:Coqblin}. e.g. for the fundamental $\Box$  of ${\rm su}(2)$ ($h=1$,${\rm N}=2$) one obtains ${\bf S}_n^2=3/4$, as appropriate for spin-1/2. 

Thus, equation (\ref{spinfermi}) supplemented by the constraint (\ref{constraint}) leads to spin operators ${\bf S}_n$ described by the Young tableau of fig.\ref{fig:young}.

\section{${\rm SU(N)}$ Thirring Model}
\label{NATM}
In this appendix we discuss the excitations, scattering matrices and form factors of the ${\rm SU(N)}$ Thirring (chiral Gross--Neveu) model. More details may be found in appendix A of Smirnov\cite{Smirnov:Form} and the literature.\cite{Koberle:Scattering,Berg:Exact,Halpern:Quantum,Kurak:Antiparticles,Andrei:Diagonalization}
\subsection{Excitations}
The excitations of the ${\rm SU(N)}$ invariant Thirring (chiral Gross--Neveu) model are ${\rm N}-1$ multiplets of fundamental particles, corresponding to the ${\rm N}-1$ fundamental representations of ${\rm SU(N)}$. Their masses are given by
\be
\label{massfun}
M_a=m\frac{\sin\pi a/{\rm N}}{\sin \pi/{\rm N}}; \quad a=1,2,\dots,{\rm N}-1,
\ee
and following Smirnov, we shall refer to the label $a$ as the ``rank'' of the particle. 
\subsection{S-Matrices}
The S-matrix describing the scattering of two (rank-1) fundamental particles in the ${\rm SU(N)}$ invariant Thirring (chiral Gross--Neveu) model is given by \cite{Koberle:Scattering,Smirnov:Form}
\be
{\mathcal S}_{\epsilon_1,\epsilon_2}^{\epsilon_1^\prime,\epsilon_2^\prime}(\theta)\equiv \ _{\epsilon_1\epsilon_2}\langle\theta_2\theta_1|{\mathcal S}^{\Box\Box}(\theta)|\theta_1\theta_2\rangle_{\epsilon_1\epsilon_2}
\ee
where $\theta=\theta_1-\theta_2$, $\epsilon\in [1,\cdots,{\rm N}]$, and the S-matrix operator acts on the two body  Hilbert space $\Box\otimes\Box$:
\be
\label{thfunfun}
{\mathcal S}^{\Box\Box}(\theta)  =   {\mathcal S}_0(\theta)\,\left(\frac{\theta {\mathcal I}-\frac{2\pi i}{{\rm N}}\,{\mathcal P}_{12}}{\theta-\frac{2\pi i}{{\rm N}}}\right).
\ee
${\mathcal I}$ and ${\mathcal P}_{12}$ are the identity and permutation operator respectively with matrix elements
\bea
\ _{\epsilon_2^\prime\epsilon_1^\prime}\langle\theta_2\theta_1|{\mathcal I}|\theta_1\theta_2\rangle_{\epsilon_1\epsilon_2} & = &\delta_{\epsilon_1}^{\epsilon_1^\prime}\delta_{\epsilon_2}^{\epsilon_2^\prime}\\
\ _{\epsilon_2^\prime\epsilon_1^\prime}\langle\theta_2\theta_1|{\mathcal P}_{12}|\theta_1\theta_2\rangle_{\epsilon_1\epsilon_2} & = & \delta_{\epsilon_1}^{\epsilon_2^\prime}\delta_{\epsilon_2}^{\epsilon_1^\prime}
\eea
and
\be
\label{s0}
{\mathcal S}_0(\theta)   =  \frac{\Gamma(1-\frac{1}{{\rm N}}+\frac{\theta}{2\pi i})\Gamma(-\frac{\theta}{2 \pi i})}{\Gamma(1-\frac{1}{{\rm N}}-\frac{\theta}{2\pi i})\Gamma(\frac{\theta}{2 \pi i})}. 
\ee
See equation (11a) of \onlinecite{Koberle:Scattering} or appendix A (p. 182) of \onlinecite{Smirnov:Form}. e.g. for ${\rm N}=2$ this reduces to equation (6) of \onlinecite{Smirnov:Form}.
Using the decompositions ${\mathcal I}={\mathcal P}^{(+)}+{\mathcal P}^{(-)}$, and ${\mathcal P}_{12}={\mathcal P}^{(+)}-{\mathcal P}^{(-)}$, one may also write (\ref{thfunfun}) in the form:
\bea
\label{thfunfun2}
{\mathcal S}^{\Box\Box}(\theta) & \equiv & \sum_{r}{\mathcal S}^{\Box\Box}_r(\theta)\, {\mathcal P}^{(r)} \nonumber \\ & = & {\mathcal S}_0(\theta)\left({\mathcal P}^{(+)}+\frac{\theta+\frac{2\pi i}{{\rm N}}}{\theta-\frac{2\pi i}{{\rm N}}}\,{\mathcal P}^{(-)}\right)
\eea
where ${\mathcal P}^{(+)}$ and ${\mathcal P}^{(-)}$ act on the symmetric and antisymmetric representations occuring in the tensor product $\Box\otimes\Box$; e.g.  $3\otimes 3=6+\bar 3$ in ${\rm SU}(3)$. Bound states correspond to poles of the S-matrix, with masses
\be 
\label{boundstatemass}
m_b  =  \sqrt{m_1^2+m_2^2+2m_1m_2\cosh(\theta_{12})}.
\ee
Since $\Gamma(z)$ is free of zeros, and exhibits simple poles at $z=0,-1,-2,\dots$,\cite{Lebedev:Special} it follows that (\ref{thfunfun2}) has a single simple pole at $\theta=2\pi i/{\rm N}$ occuring within the physical strip, $0<\theta<\pi i$. This yields the bound state mass of the second fundamental particle, $M_2 =m\sin(2\pi/{\rm N})/\sin(\pi/{\rm N})$, as given by equation (\ref{massfun}).

 The S-matrix describing the scattering of a (rank-1) fundamental particle off its conjugate (rank-${\rm N}-1$) may be obtained from (\ref{thfunfun2}) by the crossing transformation:
\be
{\mathcal S}^{\Box{\bar\Box}}(\theta)={\mathcal C}_{\Box}\, {\mathcal S}^{\Box\Box}(i\pi-\theta)\,{\mathcal C}_{\Box}
\ee
where ${\mathcal C}_{\Box}$ is the conjugation operator on $\Box$. Utilizing ${\mathcal P}^{(0)}={\mathcal C}_{\Box}\,{\mathcal P}_{12}\,{\mathcal C}_{\Box}/{\rm N}$, and ${\mathcal I}={\mathcal P}^{(\rm adj)}+{\mathcal P}^{(0)}$ one obtains 
\bea
\label{thfuncon2}
{\mathcal S}^{\Box{\bar\Box}}(\theta) & \equiv & \sum_{r}{\mathcal S}^{\Box{\bar\Box}}_r(\theta)\, {\mathcal P}^{(r)} \nonumber \\ & = & -{\mathcal S}_1(\theta) \left({\mathcal P}^{(\rm adj)}+\frac{\theta+\pi i}{\theta -\pi i}\,{\mathcal P}^{(0)}\right)
\eea
where
\be
\label{s1}
{\mathcal S}_1(\theta)=\frac{\Gamma(\frac{1}{2}+\frac{\theta}{2\pi i})}{\Gamma(\frac{1}{2}-\frac{\theta}{2\pi i})}\frac{\Gamma(\frac{1}{2}-\frac{1}{{\rm N}}-\frac{\theta}{2\pi i})}{\Gamma(\frac{1}{2}-\frac{1}{{\rm N}}+\frac{\theta}{2\pi i})},
\ee
and ${\mathcal P}^{(\rm adj)}$ and ${\mathcal P}^{(0)}$ act on the adjoint and singlet representations occuring in the tensor product $\Box\otimes{{\bar \Box}}$; e.g. $3\otimes\bar 3=8+1$ in  ${\rm SU}(3)$. In particular, for ${\rm N}=2$, (\ref{thfunfun2}) and (\ref{thfuncon2}) coincide (up to sign) as expected from the identification of $\Box$ and ${\bar\Box}$ in ${\rm SU}(2)$. Moreover, equation (\ref{thfunfun2}) is also in agreement with equation (1.6a) of reference \onlinecite{Ogievetsky:PCF}. The S-matrix (\ref{thfuncon2}) has a  pole at $\theta=\pi i-2\pi i/{\rm N}$ occuring within the physical strip, $0<\theta<\pi i$. This is a cross channel pole.

For the purpose of calculating form factors in section \ref{sect:form}, it proves useful to have the S-matrices in an integral form. Taking the logarithm of (\ref{s0}) and (\ref{s1}) and employing
\be
\ln \Gamma(z)=\int_0^\infty \frac{dt}{t}\left[(z-1)e^{-t}+\frac{e^{-tz}-e^{-t}}{1-e^{-t}}\right]
\ee 
one obtains
\be
{\mathcal S}_a(\theta)  =  \exp \left\{\int_0^\infty dx \,f_a(x)\,\sinh\left(\frac{x\theta}{\pi i}\right)\right\}
\ee
where
\bea
f_0(x) & = & \frac{2\exp(x/{\rm N})\sinh[x(1-1/{\rm N})]}{x\sinh x},\\
f_1(x) & = & \frac{2\exp(x/{\rm N})\sinh(x/{\rm N})}{x\sinh x}.
\eea
In particular, the ${\rm su}(2)$ Thirring S-matrix coincides with the sine-Gordon S-matrix with $\beta^2=8\pi$: \cite{Smirnov:Form}
\be
{\mathcal S}_a(\theta)\stackrel{{\rm N}=2}{\longrightarrow}-\exp\left\{i\int_0^\infty d\kappa \,\frac{\exp(-\pi\kappa/2)}{\kappa\cosh\pi\kappa/2}\,\sin\kappa\theta\right\}.
\ee
\subsection{Form Factors}
\label{sect:form}
In the previous paragraphs, we have discussed the elementary excitations of the ${\rm SU(N)}$ Thirring model. They are massive particles labeled by their rapdities, $\theta_i$, and carrying quantum numbers or isotopic indices, $\epsilon_i$. In order to compute correlation functions and dynamical susceptibilities, we will need the matrix elements of various physical operators, $\mathcal O$, between the vacuum and the (lowest) multiparticle excited states. Such matrix elements are termed {\em form factors}, and their computation is an important enterprise; see for example \onlinecite{Smirnov:Form}.
As is discussed in Ch. 1 of Smirnov's book,\cite{Smirnov:Form} the two-particle form factors
\be
\label{2pffdef}
F_{\mathcal O}(\theta_1,\theta_2)_{\epsilon_1,\epsilon_2}\equiv\langle 0|{\mathcal O}(0,0)|\theta_2\theta_1\rangle_{\epsilon_2,\epsilon_1}
\ee
 satisfy a matrix (Riemann--Hilbert) problem, also known as Watson's equations:
\be
F(\theta_1,\theta_2+2\pi i)_{\epsilon_1,\epsilon_2}=F(\theta_1,\theta_2)_{\epsilon_1^\prime,\epsilon_2^\prime}{\mathcal S}_{\epsilon_1,\epsilon_2}^{\epsilon_1^\prime,\epsilon_2^\prime}(\theta_{12}).
\ee
This equation may be diagonalized to yield the simpler scalar problem(s)
\be
\label{scalarrh}
F(\theta_1,\theta_2+2\pi i)=F(\theta_1,\theta_2){\mathcal S}(\theta_{12}),
\ee
where ${\mathcal S}(\theta)$ are the S-matrix eigenvalues. In particular, the Thirring current operator ${\bf J}_\mu$ (with ${\rm N}^2-1$ components) couples to the adjoint representation occuring in the tensor product $\Box\otimes\bar\Box$; the relevant eigenvalue is
\be
\label{sj}
{\mathcal S}(\theta)\equiv{\mathcal S}^{\Box\bar\Box}_{\rm adj}(\theta)=-{\mathcal S}_1(\theta). 
\ee
Another constraint on the form factors comes form Lorentz invariance. 
Under a Lorentz boost, corresponding to a simultaneous shift of all rapidities by $\Lambda$, the two-particle form factor of an operator ${\mathcal O}$ of spin $s$ satisfies:
\be
F_{\mathcal O}(\theta_1+\Lambda,\theta_2+\Lambda)=e^{s\Lambda}F_{\mathcal O}(\theta_1,\theta_2).
\ee
In particular, the left (right) component of the Thirring current has spin $s=+1$ ($s=-1$) and one obtains:
\bea
F_{j_L^a}(\theta_1,\theta_2)\propto e^{+(\theta_1+\theta_2)/2}f^{\Box{\bar\Box}}_{\rm adj}(\theta_{12}), \\
F_{j_R^a}(\theta_1,\theta_2)\propto e^{-(\theta_1+\theta_2)/2}f^{\Box{\bar\Box}}_{\rm adj}(\theta_{12}).
\eea
Note that $f^{\Box{\bar\Box}}_{\rm adj}(\theta_{12})$ is a function of $\theta_1-\theta_2$, and is thus Lorentz invariant. The form factors corresponding to the temporal and spatial components of the current may  be written:
\bea
\label{temporal}
F_{j_0^a}(\theta_1,\theta_2)\propto m\sinh\left(\frac{\theta_1+\theta_2}{2}\right)f^{\Box{\bar\Box}}_{\rm adj}(\theta_{12}), \\
\label{spatial}
F_{j_1^a}(\theta_1,\theta_2)\propto m\cosh\left(\frac{\theta_1+\theta_2}{2}\right)f^{\Box{\bar\Box}}_{\rm adj}(\theta_{12}).
\eea
Substituting (\ref{sj}) and either of (\ref{temporal}) and (\ref{spatial}) into (\ref{scalarrh}), one obtains a constraint on $f^{\Box{\bar\Box}}_{\rm adj}(\theta)$:
\be
\label{LIRH}
f^{\Box\bar\Box}_{\rm adj}(\theta-2\pi i)=f^{\Box\bar\Box}_{\rm adj}(\theta){\mathcal S}_1(\theta).
\ee
Following the general arguments of Karowski and Weisz \cite{Karowski:Exact} (equations 2.18 and 2.19) equation (\ref{LIRH}) may be solved by
\be
\label{fminint}
f^{\Box{\bar\Box}}_{\rm adj}(\theta)   =  \exp\left\{\int_0^\infty dx \, f_1(x)\,\frac{\sin^2(x\hat\theta/2\pi)}{\sinh x}\right\}
\ee
where $\hat\theta=i\pi-\theta$.\footnote{More accurately $f(\theta)$ as given by equation (\ref{fminint}) may be seen to satisfy $f(\theta-2\pi i)=f(\theta){\mathcal S}_1(\theta-2\pi i)$, as follows from equation 2.13 of ref.\cite{Karowski:Exact}; by definition ${\mathcal S}_1(\theta-2\pi i):= {\mathcal S}_1(\theta)$.} Expanding the denominator factors in powers of $e^{-2x}$, and employing the identity
\be
\exp\int_0^\infty\frac{dx}{x}\,2e^{-\beta x}\sinh\gamma x=\frac{\beta+\gamma}{\beta-\gamma},
\ee
one obtains the equivalent representation:
\bea
\label{infsimpleprod}
f^{\Box{\bar\Box}}_{\rm adj}(\theta)  & = & \prod_{l,m=0}^\infty\left[\frac{1+l+m}{1-\frac{1}{{\rm N}}+l+m}\right]^2 \times \nonumber \\  & & \hspace{-2.0cm} \left[\frac{\frac{1}{2}-\frac{1}{{\rm N}}+l+m+\frac{\theta}{2\pi i}}{\frac{1}{2}+l+m+\frac{\theta}{2\pi i}}\right]\left[\frac{\frac{3}{2}-\frac{1}{{\rm N}}+l+m-\frac{\theta}{2\pi i}}{\frac{3}{2}+l+m-\frac{\theta}{2\pi i}}\right].
\eea
Application of Euler's Formula yields:
\bea
\label{infgammaprod}
f^{\Box{\bar\Box}}_{\rm adj}(\theta)  & = &  \prod_{l=0}^\infty \left[\frac{\Gamma(1-\frac{1}{{\rm N}}+l)}{\Gamma(1+l)}\right]^2\times \nonumber \\
& & \hspace{-1.8cm} \left[\frac{\Gamma(\frac{1}{2}+l+\frac{\theta}{2\pi i})}{\Gamma(\frac{1}{2}-\frac{1}{{\rm N}}+l+\frac{\theta}{2\pi i})}\right]\left[\frac{\Gamma(\frac{3}{2}+l-\frac{\theta}{2\pi i})}{\Gamma(\frac{3}{2}-\frac{1}{{\rm N}}+l-\frac{\theta}{2\pi i})}\right]
\eea
As may be seen most clearly from (\ref{infsimpleprod}) and (\ref{infgammaprod}), this form factor is free of poles in the physical strip $0<\theta<i\pi$, and Watson's minimal equations (in Karowski--Weisz form) are explicitly satisfied: 
It is indeed, a {\it minimal} form factor. Expressions (\ref{fminint}), (\ref{infsimpleprod}) and (\ref{infgammaprod}) conform to the Karowski--Weisz normalization $F(i\pi)=1$, and have the asymptotic behaviour
\be
\lim_{\theta\rightarrow\pm\infty}f^{\Box{\bar\Box}}_{\rm adj}(\theta)\sim \exp(\pm\theta/2{\rm N}).
\ee
In the limit ${\rm N}=2$, one may write (\ref{fminint}) in the form
\bea
\label{fabform}
f^{\Box\bar\Box}_{\rm adj}(\theta)& \rightarrow & -i\sinh(\theta/2)\times \nonumber \\
& & \hspace{-1.5cm}\exp\left\{\int_0^\infty dx \frac{\sin^2(x\hat\theta/2\pi)}{x\sinh x}\left[\tanh(x/2)-1\right]\right\}
\eea
and expressions (\ref{temporal}) and (\ref{spatial}) coincide with the known results for the ${\rm SU(2)}$ invariant Thirring (or sine-Gordon) model; see equation (33) of Allen {\it et al} \cite{Allen:Fate} or let $\xi\rightarrow\infty$ in the formula for $f_\mu(\beta_1,\beta_2)$ given on page 46 of Smirnov\cite{Smirnov:Form} and note the different definition of the physical strip. In the limit ${\rm N}\rightarrow\infty$, the ${\rm SU(N)}$ Thirring model maps onto a theory of free massive fermions, as reflected in the explicit S-matrices. In this limit $f^{\Box\bar\Box}_{\rm adj}(\theta)\rightarrow 1$, and (\ref{temporal}) and (\ref{spatial}) coincide with the free fermion form factors given in equation (108) of Smirnov.\cite{Smirnov:Form}  

\bibliographystyle{/home/bhaseen/Oxford/Refs/h-physrev}

\end{document}